\documentclass[fleqn,10pt]{wlscirep}
\usepackage[utf8]{inputenc}
\usepackage[T1]{fontenc}

\usepackage{graphicx} 

\usepackage{amsmath,amsthm,amssymb,mathrsfs}
\usepackage{bm}

\usepackage{color}

\title{Spintronic reservoir computing without driving current or magnetic field}

\author[1,*]{Tomohiro Taniguchi}
\author[2]{Amon Ogihara}
\author[2]{Yasuhiro Utsumi}
\author[1,3]{Sumito Tsunegi}
\affil[1]{National Institute of Advanced Industrial Science and Technology (AIST), Research Center for Emerging Computing Technologies, Tsukuba, Ibaraki 305-8568, Japan}
\affil[2]{Department of Physics Engineering, Faculty of Engineering, Mie University, Tsu, Mie, 514-8507, Japan}
\affil[3]{PRESTO, Japan Science and Technology Agency (JST), Saitama, Japan}

\affil[*]{tomohiro-taniguchi@aist.go.jp}

\begin{abstract}
Recent studies have shown that nonlinear magnetization dynamics excited in nanostructured ferromagnets are applicable to brain-inspired computing such as physical reservoir computing. 
The previous works have utilized the magnetization dynamics driven by electric current and/or magnetic field. 
This work proposes a method to apply the magnetization dynamics driven by voltage control of magnetic anisotropy to physical reservoir computing, 
which will be preferable from the viewpoint of low-power consumption. 
The computational capabilities of benchmark tasks in single MTJ are evaluated by numerical simulation of the magnetization dynamics 
and found to be comparable to those of echo-state networks with more than 10 nodes. 
\end{abstract}

\begin{document}

\flushbottom
\maketitle
%
%


Recent development of neuromorphic computing with spintronics devices \cite{torrejon17,borders17,kudo17,grollier20}, such as pattern recognition and associative memory, has provided a bridge 
between condensed matter physics, nonlinear science, and information science, 
and become of great interest from both fundamental and practical viewpoints. 
In particular, an application of nonlinear magnetization dynamics in ferromagnets to physical reservoir computing 
\cite{maas02,jaeger04,verstraeten07,hermans10,appeltant11,paquot12,brunner13,nakajima15,nakayama16,sande17,fujii17,dion18,nakajima20,goto21,nakajima21} 
is an exciting topic \cite{torrejon17,furuta18,tsunegi18,bourianoff18,nakane18,markovic19,tsunegi19,riou19,nomura19,yamaguchi20,yamaguchi20srep,akashi20}. 
Physical reservoir computing is a kind of recurrent neural network, which has recurrent interaction among large number of neurons in artificial neural network 
and, for example, recognizes a time sequence of the input data, such as human voice and movie, from the dynamical response in nonlinear physical systems \cite{nakajima21}. 
In reservoir computing, only the weights between neurons and output are trained, whereas the weights among neurons are randomly given and fixed, and therefore, low calculation cost of training is expected. 
It has been shown that several kinds of physical systems, 
such as optical circuit \cite{paquot12}, soft matter \cite{nakajima15}, quantum matter \cite{fujii17}, fluid \cite{goto21}, and spintronics devices, can be used as reservoir for information processing \cite{nakajima21}. 


In physical reservoir computing with spintronics devices, nonlinear magnetization dynamics has been excited in nanostructured ferromagnets by applying electric current and/or magnetic field. 
For example, spin-transfer effect \cite{slonczewski96,berger96} has been frequently used to excite an auto-oscillation of the magnetization in magnetic tunnel junctions (MTJs)
\cite{torrejon17,furuta18,tsunegi18,bourianoff18,markovic19,tsunegi19,riou19,yamaguchi20,yamaguchi20srep,akashi20}, 
where the spin angular momentum from conducting electrons carrying electric current is transferred to ferromagnet and excites magnetization dynamics. 
It is, however, preferable to excite magnetization dynamics without driving current and magnetic field 
from the viewpoints of low-power consumption and simple implementation. 


In this work, we propose that physical reservoir computing can be performed by magnetization dynamics induced by voltage control of magnetic anisotropy in solid devices 
\cite{weisheit07,duan08,maruyama09,nakamura09,tsujikawa09,shiota09,nozaki10,shiota11,wang11,shiota12,kanai12,grezes16,nozaki17,miwa17,okada18,sugihara19,yamamoto19,nozaki20}. 
The voltage control of magnetic anisotropy is a fascinating technology as the low-power information writing scheme in magnetoresistive random access memory, instead of using spin-transfer torque effect. 
An application of electric voltage to a metallic ferromagnet/insulator interface modifies electron states near the interface \cite{duan08,nakamura09,tsujikawa09} and/or induces magnetic moment \cite{miwa17}, 
and changes magnetic anisotropy. 
The magnetization in the ferromagnetic metal changes its direction to minimize the magnetic anisotropy energy. 
Therefore, the voltage application can cause the relaxation dynamics of the magnetization in the ferromagnet. 
In the practical application of nonvolatile random access memory, an external magnetic field is necessary to achieve a deterministic magnetization switching guaranteeing reliable writing \cite{shiota11,wang11,shiota12,kanai12}. 
On the other hand, we notice that the magnetization switching, as well as magnetic field, is not a necessary condition in physical reservoir computing. 
Accordingly, the voltage control of magnetic anisotropy can be used to realize physical reservoir computing by spintronics devices without driving current or magnetic field. 
Here, we perform numerical simulation of the Landau-Lifshitz-Gilbert (LLG) equation 
and find that the computational capabilities of benchmark tasks in single spintronics device are comparable to those of echo-state networks with more than 10 nodes. 


\section*{Model}


\begin{figure}
\centerline{\includegraphics[width=1.0\columnwidth]{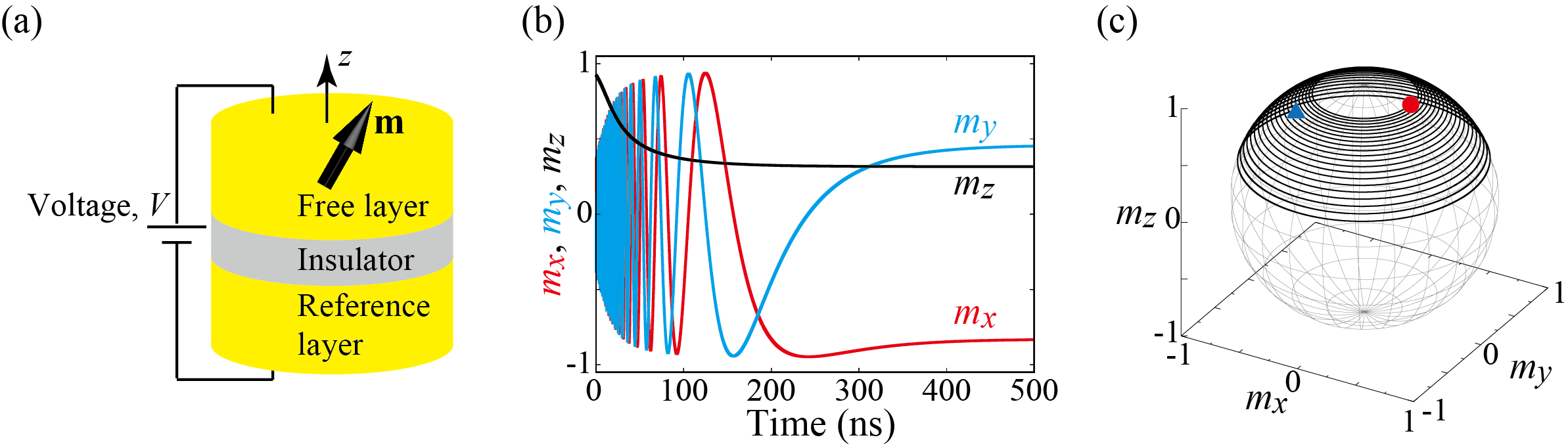}}
\caption{
         (a) Schematic illustration of an MTJ. 
             The unit vector pointing to the magnetization direction in the ferromagnetic free layer is $\mathbf{m}$. 
             The $z$ axis is perpendicular to the film plane. 
         (b) An example of the time evolutions of $m_{x}$ (red), $m_{y}$ (blue), and $m_{z}$ (black). 
         (c) Trajectory of the relaxation dynamics on a sphere. 
         In (b) and (c), the first order magnetic anisotropy field $H_{\rm K1}$ is changed from $-0.1 H_{\rm K2}$ to $-0.9 H_{\rm K2}$ by the voltage application. 
         The red circle and blue triangle in (c) represent the initial and final states of the dynamics. 
         \vspace{-3ex}}
\label{fig:fig1}
\end{figure}



\subsection*{LLG equation}

The system under investigation is a cylinder-shaped MTJ schematically shown in Fig. \ref{fig:fig1}(a), 
where the $z$ axis is perpendicular to the film plane. 
The MTJ consists of ferromagnetic free layer, MgO insulator, and ferromagnetic reference layer. 
The ferromagnetic free layer has the perpendicular magnetic anisotropy, where the magnetic energy density is given by 
\begin{equation}
  \varepsilon
  =
  \sum_{i=x,y,z}
  2\pi M^{2} N_{i}
  m_{i}^{2}
  +
  K_{1}
  \left(
    1
    -
    m_{z}^{2}
  \right)
  +
  K_{2}
  \left(
    1
    -
    m_{z}^{2}
  \right)^{2}. 
  \label{eq:energy}
\end{equation}
The first term on the right-hand side in Eq. (\ref{eq:energy}) represents the shape magnetic anisotropy energy density with the saturation magnetization $M$ and the demagnetization coefficients $N_{i}$. 
Since we assume the cylinder shape, $N_{x}=N_{y}$. 
The unit vector pointing in the magnetization direction of the free layer is denoted as $\mathbf{m}=(m_{x},m_{y},m_{z})$. 
The second and third terms are the first and second order magnetic anisotropy energy densities with the coefficients $K_{1}$ and $K_{2}$. 
Note that the energy density relates to the magnetic field inside the free layer as 
\begin{equation}
  \mathbf{H}
  =
  \left[
    H_{\rm K1}
    +
    H_{\rm K2}
    \left(
      1
      -
      m_{z}^{2}
    \right)
  \right]
  m_{z}
  \mathbf{e}_{z}, 
  \label{eq:magnetic_field}
\end{equation}
where $H_{\rm K1}=(2K_{1}/M)-4\pi M (N_{z}-N_{x})$ and $H_{\rm K2}=4K_{2}/M$; see also Methods. 
The magnetization in the reference layer points to the $z$ direction, and therefore, $m_{z}$ is experimentally measured through tunnel magnetoresistance effect. 


The first order magnetic anisotropy energy coefficient $K_{1}$ consists of the bulk and interfacial contributions, $K_{\rm v}$ and $K_{\rm i}$, and the voltage-controlled magnetic anisotropy effect described as 
$K_{1}d=K_{\rm v}d+K_{\rm i}-\eta \mathscr{E}$. 
The thickness of the ferromagnetic free layer is $d$, 
whereas $\mathscr{E}=V/d_{\rm I}$ is the electric field with the voltage $V$ and the thickness of the insulator $d_{\rm I}$. 
In typical MTJs consisting of CoFeB free layer and MgO insulator, $K_{\rm i}$ dominates in $K_{1}$, 
where $K_{\rm i}$ increases with the increase of the composition of Fe \cite{yakata09,ikeda10,kubota12}. 
It can reach on the order of 1.0 mJ/m${}^{2}$ at maximum, which in terms of magnetic field, $2K_{\rm i}/(Md)$, is typically on the order of 1 T. 
Note that the magnitude of the shape magnetic anisotropy field $-4\pi M (N_{z}-N_{x})$ is also on the order of 1 T, 
where a typical value of the saturation magnetization in CoFeB, i.e., $M$ of about $1000$ emu/c.c., is assumed. 
As a result of the competition between them, 
the ferromagnetic free layer in the absence of the voltage application can be either in-plane or perpendicular-to-plane magnetized \cite{yakata09,ikeda10,kubota12}. 
The voltage control of magnetic anisotropy also modifies the magnetic anisotropy field $H_{\rm K1}$ 
through the modification of the electron occupation states near the ferromagnetic interface \cite{duan08,nakamura09,tsujikawa09} 
and/or the generation of the magnetic dipole moment \cite{miwa17}. 
The coefficient of the voltage-controlled magnetic anisotropy effect, $\eta$, is recently achieved in the experiment to be about 300 fJ/(Vm) \cite{nozaki17,nozaki20}, 
whereas the thickness of the insulator is about $2.5$ nm. 
A typical values of the applied voltage is about $0.5$ V at maximum \cite{sugihara19}. 
Thus, the tunable range of the magnetic anisotropy by the voltage application in terms of the magnetic field, $(2 |\eta| V)/(Mdd_{\rm I})$, is about 1.0 kOe, 
where we assume that $M=1000$ emu/c.c., $d=1$ nm, $d_{\rm I}=2.5$ nm, and $|\eta|=250$ fJ/(V m). 
Note that the sign of the voltage-controlled magnetic anisotropy effect depends on that of the voltage. 
Summarizing these contributions, $H_{\rm K1}$ in the presence of the voltage can also be either positive or negative, 
depending on the materials and their compositions, as well as the magnitude and sign of the applied voltage. 
For example, Ref. \cite{shiota09} uses an in-plane magnetized ferromagnet, i.e., $H_{\rm K1}<0$ for $V=0$. 
The voltage control of magnetic anisotropy in Ref. \cite{shiota09} enhances the perpendicular anisotropy $K_{1}$ and makes $H_{\rm K1}$ positive at nonzero $V$. 
On the other hand, perpendicularly magnetized free layers where $H_{\rm K1}>0$ for $V=0$ have been used in Ref. \cite{kanai12}. 
Contrary to $H_{\rm K1}$, the dependence of $H_{\rm K2} \propto K_{2}$ on the applied voltage is still unclear, 
where Ref. \cite{okada18} reports that $H_{\rm K2}$ is approximately independent of the voltage 
while Ref. \cite{sugihara19} observes the voltage dependence of $H_{\rm K2}$. 
Throughout this paper, for simplicity, we assume that only $H_{\rm K1}$ depends on the voltage. 
As mentioned in the following, we performed numerical simulation by changing the value of $H_{\rm K1}$. 
It means that we do not specify the size (the thickness and cross-section area) of MTJ explicitly 
because $H_{\rm K1}$ includes the information of the shape of MTJ through the demagnetization coefficients $N_{i}$. 
It is, however, useful to mention that macrospin model has been proven to work well to describe the magnetization dynamics for MTJ 
whose typical size is $1$-$2$ nm in thickness and the diameter less than 200 nm \cite{shiota11,shiota12,yamamoto19}. 


In typical experiments on voltage control of magnetic anisotropy, a relatively thick (typically $1.5$-$2.5$ nm) MgO barrier is used as an insulator \cite{shiota12,kanai12,yamamoto19}, 
compared with MTJ manipulated by spin-transfer torque, where the thickness of the barrier is about $1.0$ nm \cite{yakushiji13}. 
As a result, the resistance of MTJ used for experiments of voltage control of magnetic anisotropy, on the order of $10$-$100$ k$\Omega$, 
is two or three orders of magnitude larger than that used for spin-transfer torque experiments. 
On the other hand, the maximum voltage used in both experiments is almost identical. 
Accordingly, current flowing in MTJ used for experiments of voltage control of magnetic anisotropy is two or three orders of magnitude smaller than that used for spin-transfer torque experiments 
(see also Methods). 
In this sense, we mention that the driving force of magnetization dynamics is voltage control of magnetic anisotropy effect, although current cannot be completely zero in experiments. 
As mentioned in Methods, typical value of current $I$ flowing in MTJ is on the order of $1$ $\mu$A, while the current used in physical reservoir computing utilizing spin-transfer torque is on the order of $1$ mA \cite{yamaguchi20srep}. 
On the other hand, the magnitude of the voltage $V$ applied to MTJ is nearly the same for both experiments on voltage control of magnetic anisotropy and spin-transfer effects. 
Accordingly, using the voltage control of magnetic anisotropy effect could reduce energy consumption by three orders. 


The magnetization in equilibrium points to the direction at which the energy density is minimized. 
For example, when $H_{\rm K1}>(<)0$ and $H_{\rm K2} = 0$, the energy is minimized when the magnetization is parallel (perpendicular) to the $z$ axis. 
Another example is studied in Ref. \cite{matsumoto15}, where, if $H_{\rm K1}<0$ and $|H_{\rm K1}|<H_{\rm K2}$, 
the energy density $\varepsilon$ is minimized when $m_{z}=\pm\sqrt{1-(|H_{\rm K1}|/H_{\rm K2})}$. 
When the voltage is applied to the MTJ and the minimum energy state is changed as a result, the magnetization relaxes to the state. 
The relaxation dynamics is described by the LLG equation, 
\begin{equation}
  \frac{d \mathbf{m}}{dt}
  =
  -\gamma
  \mathbf{m}
  \times
  \mathbf{H}
  +
  \alpha
  \mathbf{m}
  \times
  \frac{d \mathbf{m}}{dt},
  \label{eq:LLG}
\end{equation}
where $\gamma$ and $\alpha$ are the gyromagnetic ratio and the Gilbert damping constant, respectively. 
Note that the macrospin model works well to describe the magnetization dynamics driven by the voltage application \cite{shiota11,shiota12,grezes16}. 
The values of the parameters used in the following are derived from typical experiments 
\cite{maruyama09,shiota09,nozaki10,shiota11,wang11,shiota12,kanai12,grezes16,okada18,sugihara19}. 
The gyromagnetic ratio and the Gilbert damping constant are $\gamma=1.764 \times 10^{7}$ rad/(Oe s) and $\alpha=0.01$. 
The second order magnetic anisotropy field $H_{\rm K2}$ is 500 Oe. 


Let us show an example of the magnetization dynamics driven by the voltage control of magnetic anisotropy. 
We firstly set $H_{\rm K1}$ to be $H_{\rm K1}^{(0)}=-0.1 H_{\rm K2}=-50$ Oe and solve the LLG equation with an arbitrary initial condition. 
The magnetization saturates to a certain point where $m_{z}$ saturates to $m_{z}\to m_{z}^{(0)}\simeq 0.95$. 
We use this state as a new initial state and solve the LLG equation by changing $H_{\rm K1}$ to $H_{\rm K1}^{(1)}=-0.9 H_{\rm K2}=-450$ Oe. 
Then, the magnetization starts to move to a new stable state where $m_{z}$ saturates to $m_{z}\to m_{z}^{(1)} \simeq 0.32$. 
Figures \ref{fig:fig1}(b) and \ref{fig:fig1}(c) show time evolution of $\mathbf{m}$ and its spatial orbit from the initial state of $m_{z}^{(0)}$ to the final state $m_{z}^{(1)}$. 
We confirm that the initial and final states are those expected from the minimum energy state mentioned above, 
i.e., $m_{z}^{(0)}=\sqrt{1-|H_{\rm K1}^{(0)}|/H_{\rm K2}}=\sqrt{1-0.1}\simeq 0.95$ and $m_{z}^{(1)}=\sqrt{1-|H_{\rm K1}^{(1)}|/H_{\rm K2}}=\sqrt{1-0.9}\simeq 0.32$. 
We emphasize that $m_{z}$ monotonically changes with respect to the change of $H_{\rm K1}$. 
Since the value of $H_{\rm K1}$ can be manipulated by the voltage application, 
the time evolution of $m_{z}$ can be used to identify the value of the applied voltage. 
The estimation of the input data, which is the sequence of the applied voltage in the present case, from the dynamical response of physical system is the aim of physical reservoir computing. 
Therefore, the magnetization dynamics driven by the voltage control of magnetic anisotropy is applicable to physical reservoir computing. 
In the following, we evaluate its computational ability. 


\section*{Results}


\subsection*{Memory capacity}

The ability in physical system for reservoir computing has been quantified by memory capacity \cite{fujii17,goto21,furuta18,tsunegi18,tsunegi19,yamaguchi20,akashi20}. 
The memory capacity corresponds to the number of past data physical reservoir can store. 
%
%
For example, let us imagine injecting random binary input $b=0$ or $1$ to reservoir, as done in experiments \cite{tsunegi18,tsunegi19}. 
The input data are often injected as pulses with the pulse width of $t_{\rm p}$, i.e., the value of $b$ is constant during time $t_{\rm p}$. 
Therefore, it is convenient to add a suffix $k=1,2,\cdots$ to $b$ as $b_{k}$ to distinguish the order of the input data. 
We also introduce an integer $D=0,1,2,\cdots$, called delay, characterizing the order of the past input data. 
In this case, 
\begin{equation}
  y_{k,D}^{\rm STM}
  =
  b_{k-D},
  \label{eq:target_STM}
\end{equation}
are called target data of short-term memory (STM) task. 
We predict the value of the target data from the output of the reservoir and evaluate the reproducibility. 
The predicted data are called system output. 
The reproducibility is quantified by the correlation coefficient between the target data and system output. 
Roughly speaking, if the reservoir can reproduce the past data up to $D$, the STM capacity is defined as $D$. 
There is another kind of memory capacity, called parity-check (PC) capacity, where the target data are defined as 
\begin{equation}
  y_{k,D}^{\rm PC}
  =
  \sum_{\ell=0}^{D}
  b_{k-D+\ell}\ 
  \left(
    {\rm mod}2
  \right). 
  \label{eq:target_PC}
\end{equation}
According to their definitions, the STM and PC capacities quantify the number of the target data the reservoir can store, 
where the target data are defined as linear and nonlinear transformations of the input data, respectively. 
A large memory capacity means that reservoir can store, recognize, and/or predict large data. 
See also Methods for the detail of the evaluation method of these capacities. 

\begin{figure}
\centerline{\includegraphics[width=1.0\columnwidth]{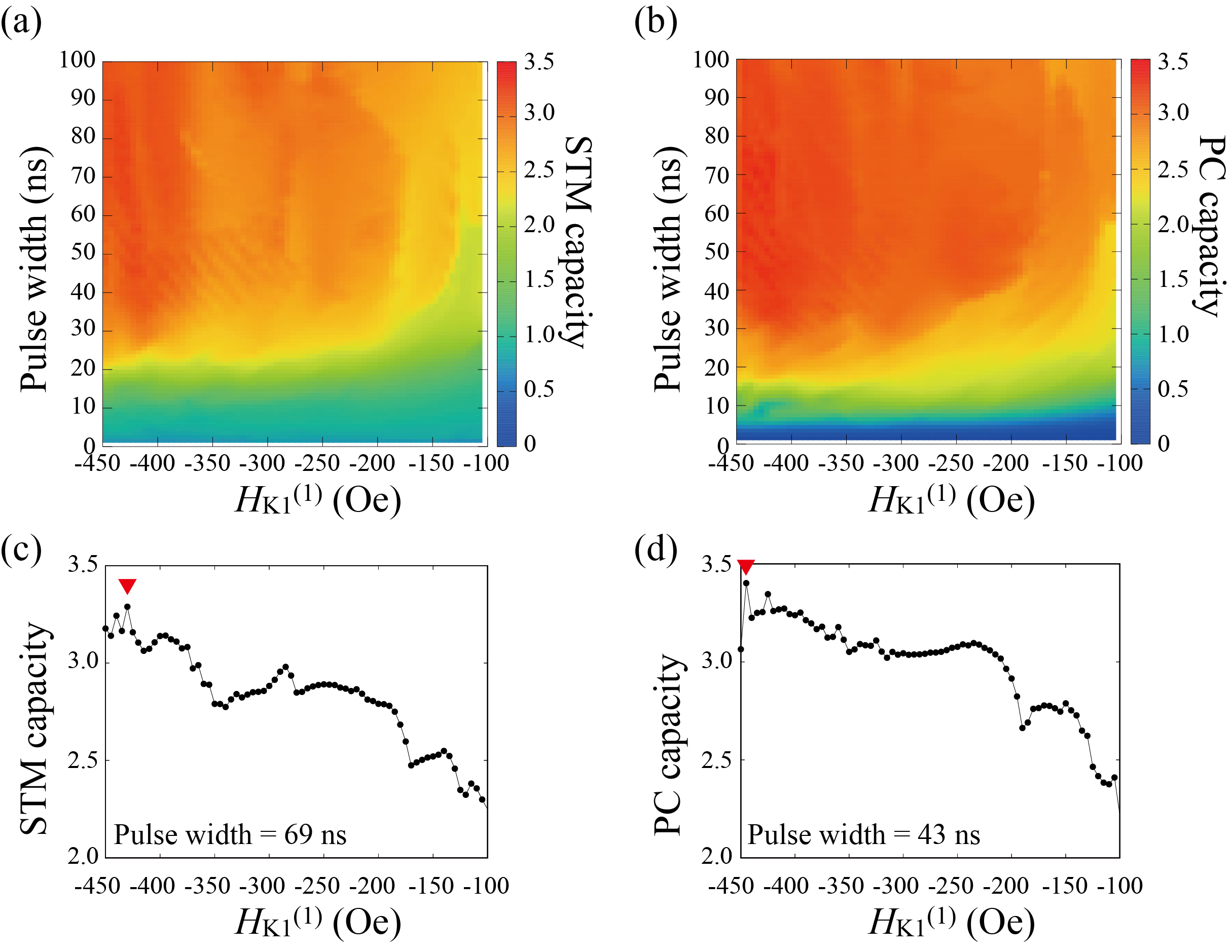}}
\caption{
         Dependences of (a) STM (linear) and (b) PC (nonlinear) capacities on the pulse width and the first order magnetic anisotropy field. 
         Their highest values are indicated by the red triangles in (c) and (d). 
         \vspace{-3ex}}
\label{fig:fig2}
\end{figure}

In the present system, the random binary inputs are injected as voltage pulses, which change the first order magnetic anisotropy field $H_{\rm K1}$ as 
\begin{equation}
  H_{\rm K1}
  =
  H_{\rm K1}^{(0)}
  \left(
    1
    -
    b_{k} 
  \right)
  +
  H_{\rm K1}^{(1)}
  b_{k}. 
  \label{eq:input_STM_PC}
\end{equation}
Accordingly, when the input is $b_{k}=0$ ($1$), the value of $H_{\rm K1}$ is $H_{\rm K1}^{(0)}$ [$H_{\rm K1}^{(1)}$].
In the following, we fix $H_{\rm K1}^{(0)}=-50$ Oe, i.e., $H_{\rm K1}^{(0)}/H_{\rm K2}=-0.1$, 
whereas $H_{\rm K1}^{(1)}$ varies in the range of $-450 \le H_{\rm K1}^{(1)} \le -100$ Oe, i.e., $-0.9 \le H_{\rm K1}^{(1)}/H_{\rm K2} \le -0.2$. 
Figures \ref{fig:fig2}(a) and \ref{fig:fig2}(b) show the STM and PC capacities as a function of $H_{\rm K1}^{(1)}$ and the pulse width of the input data. 
The highest value of the STM capacity, $3.29$, is found at the conditions of $H_{\rm K1}^{(1)}=-430$ Oe and $t_{\rm p}=69$ ns, as shown in Fig. \ref{fig:fig2}(c). 
On the other hand, the highest value of the PC capacity, $3.40$, is found at the conditions of $H_{\rm K1}^{(1)}=-445$ Oe and $t_{\rm p}=43$ ns, as shown in Fig. \ref{fig:fig2}(d). 



\begin{figure}
\centerline{\includegraphics[width=1.0\columnwidth]{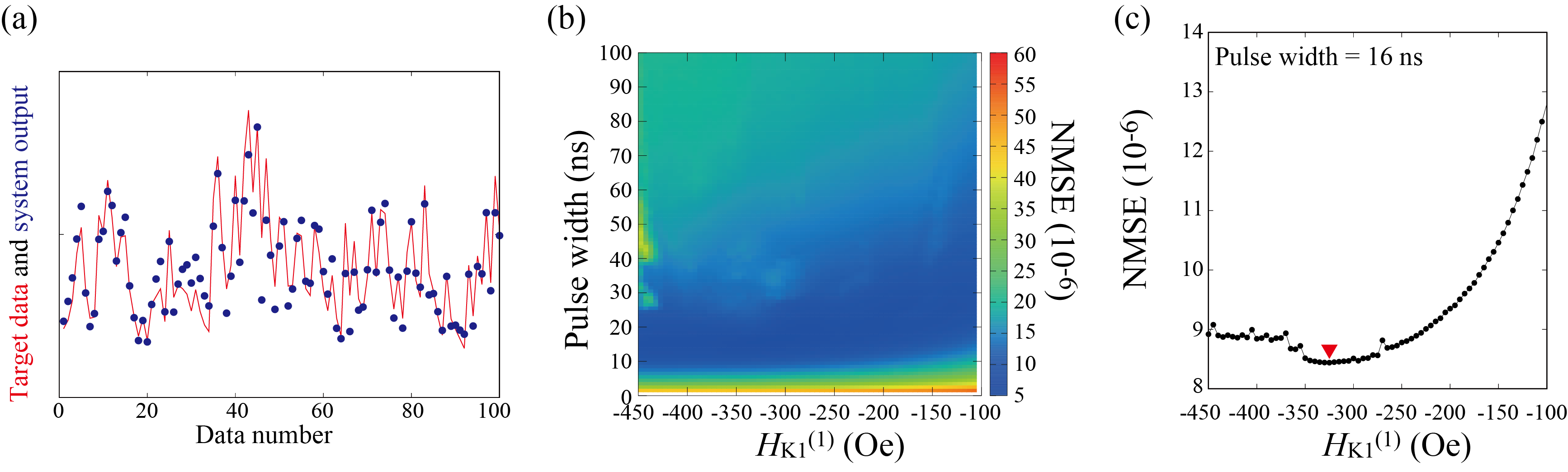}}
\caption{
         (a) Examples of the target data (red line) and the system output (blue dots) of NARMA2 task, where $t_{\rm p}=16$ ns and $H_{\rm K1}^{(1)}=-325$ Oe. 
         (b) Dependence of the NMSE of NARMA2 task on the pulse width and the first order magnetic anisotropy field. 
         The lowest value of the NMSE is indicated by the red triangle in (c). 
         \vspace{-3ex}}
\label{fig:fig3}
\end{figure}




\subsection*{NARMA task}

Another benchmark task to quantify the computational ability of physical system to reservoir computing is nonlinear autoregressive moving average (NARMA) task 
\cite{nakajima15,fujii17,goto21,akashi20,atiya00}. 
NARMA task is a function-approximation task to reproduce a nonlinear function defined from input data by using output data in recurrent neural networks. 
The task is classified as NARMA$D$ with $D=2,5,10$ and so on, where $D$ represents the delay included in the nonlinear function. 
In other words, the target data of NARMA$D$ task consist of data defined until $D$ times before from the present data. 
For example, in NARMA2 task, the system is aimed to reproduce the target data, 
\begin{equation}
  y_{k+1}^{\rm NARMA2}
  =
  0.4 y_{k}^{\rm NARMA2}
  +
  0.4 
  y_{k}^{\rm NARMA2} 
  y_{k-1}^{\rm NARMA2}
  +
  0.1 
  z_{k}^{3}
  +
  0.1,
  \label{eq:NARMA2_def}
\end{equation}
from output data, where $z_{k}=0.2 r_{k}$ is defined from uniform random input data $r_{k}$ at a discrete time $k$; see Methods. 
The computational ability of NARMA task is evaluated from normalized mean-square error (NMSE) defined as 
\begin{equation}
  {\rm NMSE}
  =
  \frac{ \sum_{k} \left( y_{k}^{\rm NARMA2} - v_{k}^{\rm NARMA2} \right)^{2}}{\sum_{k} \left( y_{k}^{\rm NARMA2} \right)^{2}}, 
  \label{eq:NMSE_def}
\end{equation}
where $v_{k}^{\rm NARMA2}$ is the data reproduced from the output data (see also Methods). 
A low NMSE corresponds to high reproducibility of the target data. 
Figure \ref{fig:fig3}(a) shows an example of the target data (red line), $y_{k}^{\rm NARMA2}$, and the system output (blue dots). 
By evaluating the difference between the target data and the system output as such, the NMSE is obtained as shown in Fig. \ref{fig:fig3}(b). 
The NMSE is on the order of $10^{-6}-10^{-5}$ and is minimized to be $8.43\times 10^{-6}$ at $t_{\rm p}=16$ ns and $H_{\rm K1}^{(1)}=-325$ Oe; see Fig. \ref{fig:fig3}(c). 


\section*{Discussion}

We have developed theoretical analysis of the magnetization dynamics in nanostructured ferromagnetic multilayers driven by the voltage control of magnetic anisotropy, 
and showed that the dynamics is applicable to physical reservoir computing through the evaluations of the memory capacity and the NMSE of NARMA task. 
Neither electric current nor external magnetic field is introduced in the computation, 
contrary to the previous works focusing on the application to nonvolatile memory, because magnetization switching is unnecessary. 
This fact will be preferable for reducing power consumption in reservoir computing. 


Figures \ref{fig:fig2}(a) and \ref{fig:fig2}(b) show that the memory capacity increases with the difference between $H_{\rm K1}^{(0)}$ and $H_{\rm K1}^{(1)}$ increasing. 
This is because when the difference between $H_{\rm K1}^{(0)}$ and $H_{\rm K1}^{(1)}$ is large, the range of the dynamical response of $m_{z}$ also becomes large, 
which makes it easy to identify the input data from the change of $m_{z}$. 
Due to a similar reason, the memory capacity increases with the increase of the pulse width. 
When the pulse width is relatively long, the change of $m_{z}$ during a pulse injection becomes large, which again makes it easy to identify the input data. 
However, when the pulse width is sufficiently long, $m_{z}$ finally saturates to a stable state, and becomes approximately constant, as implied from Fig. \ref{fig:fig1}(b). 
When $m_{z}$ becomes constant, it becomes impossible to estimate the past input from the present output. 
Therefore, the memory capacity does not increase monotonically with the pulse width increasing. 
As written above, the STM and PC capacities are maximized at the pulse width of $69$ and $43$ ns, respectively. 
A similar trend is found in NARMA2 task, where low NMSEs are achieved in a relatively large $H_{\rm K1}^{(1)}$ region. 
Note that the memory capacity at the maximum was found to be about $3$, which is comparable to the computational ability of echo-state network with approximately 10 nodes \cite{furuta18,yamaguchi20}. 
The value is also comparable or larger than that obtained by the other single spintronics reservoirs without additional circuits \cite{furuta18,tsunegi18,yamaguchi20srep}, 
driven by electric current and/or magnetic field. 
This might be due to a matching between the relaxation time of the output signal and the pulse width. 
Another possible reason is a large change in the dynamical amplitude, compared with an oscillator system \cite{yamaguchi20srep}. 
The NMSE of NARMA2 task, minimized to be on the order of $10^{-6}$, is also comparable or lower than that found in soft robot \cite{nakajima15} and echo-state network with nodes more than 10 \cite{goto21}. 
These results indicate the potential applicability of an MTJ driven by the voltage-controlled magnetic anisotropy effect to physical reservoir computing. 


An empirical rule shared among the research community is that the computational ability of physical reservoir computing is maximized at the edge of chaos \cite{nakayama16,akashi20,bertschinger04}. 
Simultaneously, an existence of chaos might lose the reproducibility of the computation due to the sensitivity to initial states. 
Note that chaos is prohibited in the present system when random inputs are absent. 
This is because the magnetization dynamics are described by two variables, $\theta=\cos^{-1}m_{z}$ and $\varphi=\tan^{-1}(m_{y}/m_{x})$, 
whereas the Poincar\'e-Bendixson theorem argues that chaos does not appear in a two dimensional system. 
When the random input are injected to the MTJ, the system becomes nonautonomous due to the presence of time-dependent torque. 
In this case, the number of the dimension in the phase space becomes three, and the possibility to induce chaos becomes finite. 
For example, Ref. \cite{akashi20} reported the appearance of chaos in a spin-torque oscillator due to the injection of random input current. 
However, we should notice that the presence of time-dependent input does not necessarily guarantee the presence of chaos. 
The identification of chaos is done by, for example, evaluating the Lyapunov exponent. 
The Lyapunov exponent quantifies the time evolution of an infinitesimal difference given at the initial state. 
The positive Lyapunov exponent implies the presence of chaos. 
On the other hand, when the Lyapunov exponent is negative, the dynamics saturate to fixed points. 
When the Lyapunov exponent is zero, the dynamics is periodic. 
The dynamics with negative or zero Lyapunov exponent are classified as ordered dynamics. 
Since the LLG equation describes the relaxation dynamics to stable states, one might consider that the largest Lyapunov exponent of an MTJ in the absence of random inputs is negative. 
However, notice that the axial symmetry of the present system enables us to move the magnetization rotating around the $z$ axis without energy injection. 
In fact, the energy density, as well as the equation of motion for $m_{z}$ depends on $m_{z}$ only, as explained in Methods; 
in other words, the equation of motion for $\theta$ is independent of $\varphi$. 
As a result, an infinitesimal difference given to the phase $\varphi$ is not shortened by the LLG equation. 
Therefore, the largest Lyapunov exponent in the absence of the random input is zero. 
The fact that the equation of motion for $\theta$ depends on $\theta$ only also implies the absence of homoclinic bifurcations, as well as chaos, 
even when the pulse data, independent of $\theta$ and $\varphi$, are injected; 
in fact, the numerically evaluated Lyapunov exponent was zero, as explained in Methods. 
The absence of chaos indicates the reproducibility of the computation in the present reservoir.



In summary, we perform numerical experiments of the magnetization dynamics in an MTJ driven by the voltage control of magnetic anisotropy. 
Injecting the voltage pulse to the MTJ, the magnetization changes its direction to minimize the magnetic anisotropy energy. 
The time evolution of the relaxation dynamics reflects the value of the input voltage, and therefore, can be used to reproduce the time sequence of the input data. 
We evaluate the computing abilities, such as the memory capacity and the error in the reproducibility, of common benchmark task, 
and show that even a single MTJ can show high computing performance comparable to echo-state network consisting of multiple nodes more than 10. 
Since neither electric current nor external magnetic field is necessary, the proposal here will be of interest for energy-saving computing technologies. 


\section*{Methods}


\subsection*{Definition of magnetic field and relaxation time}

The magnetic field $\mathbf{H}$ relates to the energy density $\varepsilon$ as $\mathbf{H}=-\partial \varepsilon/\partial (M \mathbf{m})$, 
and therefore, is obtained from Eq. (\ref{eq:energy}) as 
\begin{equation}
  \mathbf{H}
  =
  \begin{pmatrix}
    -4\pi M N_{x} m_{x} \\
    -4\pi M N_{y} m_{y} \\
    \left[ \left( 2K_{1}/M \right)-4\pi M N_{z} \right] m_{z} + \left( 4K_{2}/M \right) \left( 1-m_{z}^{2} \right) m_{z}
  \end{pmatrix}.
  \label{eq:field_orig}
\end{equation}
We should note that the magnetization dynamics described by the LLG equation is unchanged by adding a term proportional to $\mathbf{m}$ to $\mathbf{H}$ 
because the LLG equation conserves the magnitude of $\mathbf{m}$. 
Adding a term as such corresponds to shifting the origin of the energy density $\varepsilon$ by a constant. 
In the present case, we added a term $4\pi M N_{x}\mathbf{m}$ to $\mathbf{H}$ and obtained Eq. (\ref{eq:magnetic_field}), 
where we should remind that $N_{x}=N_{y}$ because we assume a cylinder shaped MTJ. 
The added term to $\mathbf{H}$ shifts the origin of the energy density $\varepsilon$ by the constant $-2\pi M^{2} N_{x} \mathbf{m}^{2}=-2\pi M^{2}N_{x}$ and makes it depend on $m_{z}$ only. 


The LLG equation in the present system can be integrated as 
\begin{equation}
  t
  =
  \frac{1+\alpha^{2}}{\alpha \gamma \left(H_{\rm K1}+H_{\rm K2}\right)}
  \left[
    \log
    \left(
      \frac{\cos\theta_{\rm f}}{\cos\theta_{\rm i}}
    \right)
    -
    \frac{H_{\rm K1}+H_{\rm K2}}{H_{\rm K1}}
    \log
    \left(
      \frac{\sin\theta_{\rm f}}{\sin\theta_{\rm i}}
    \right)
    +
    \frac{H_{\rm K2}}{2H_{\rm K1}}
    \log
    \left(
      \frac{H_{\rm K1}+H_{\rm K2} \sin^{2}\theta_{\rm f}}{H_{\rm K1}+H_{\rm K2} \sin^{2}\theta_{\rm i}}
    \right)
  \right], 
  \label{eq:relaxation_time}
\end{equation}
where $\theta_{\rm i}$ and $\theta_{\rm f}$ are the initial and final values of $\theta=\cos^{-1}m_{z}$. 
Equation (\ref{eq:relaxation_time}) provides the relaxation time from $\theta=\theta_{\rm i}$ to $\theta=\theta_{\rm f}$. 
Note that the relaxation time is scaled by $\alpha \gamma H_{\rm K1}/(1+\alpha^{2})$ and $H_{\rm K2}/H_{\rm K1}$, which can be manipulated by the voltage control of magnetic anisotropy. 
We also note that Eq. (\ref{eq:relaxation_time}) has logarithmic divergence due to asymptotic behavior in the relaxation dynamics.

\subsection*{Role of spin-transfer torque}

\begin{figure}
\centerline{\includegraphics[width=1.0\columnwidth]{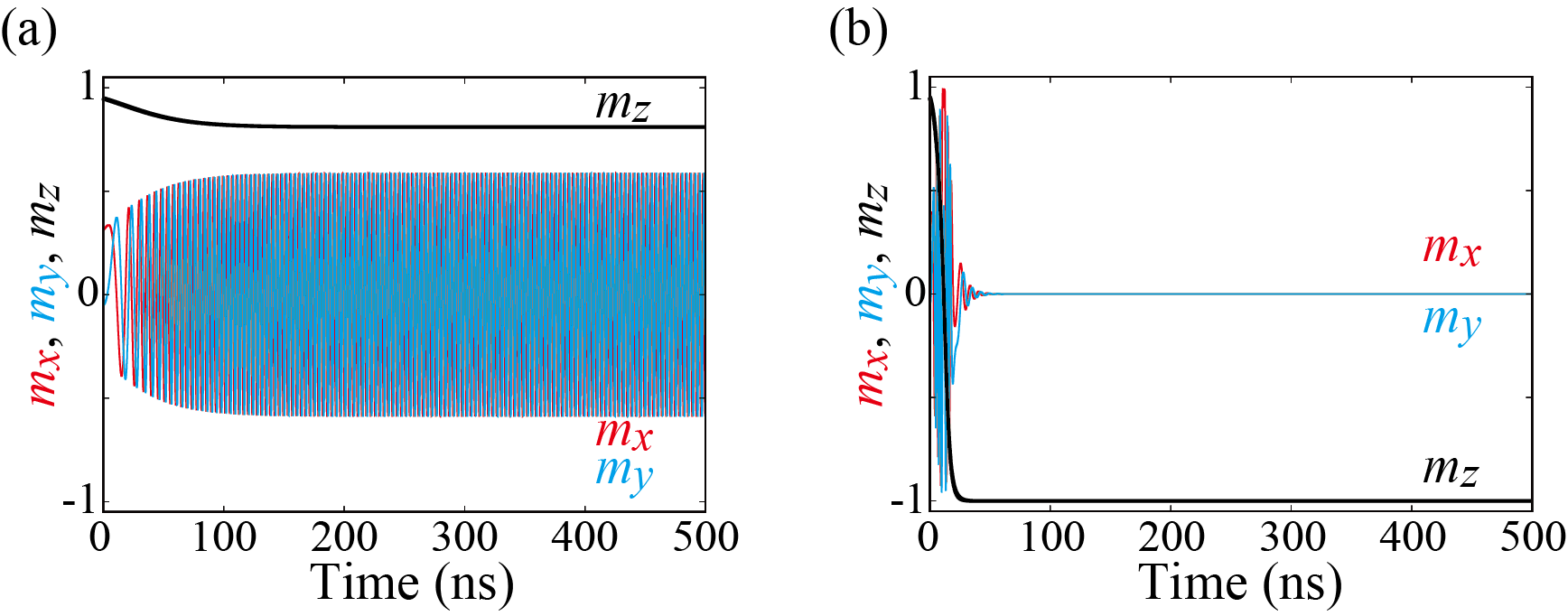}}
\caption{
        Examples of the time evolutions of $m_{x}$ (red), $m_{y}$ (blue), and $m_{z}$ (black) in the presence of spin-transfer torque, 
        where the current density is (a) $0.06$ MA/cm${}^{2}$ and (b) $0.6$ MA/cm${}^{2}$. 
         \vspace{-3ex}}
\label{fig:fig4}
\end{figure}

We neglected spin-trasnfer torque in the main text because the current magnitude in typical MTJ used for voltage control of magnetic anisotropy effect is usually small. 
For example, when using typical values \cite{okada18,yamamoto19} for the voltage ($0.4$ V), resistance ($60$ k$\Omega$), and cross-section being $\pi \times 60^{2}$ nm${}^{2}$, 
the value of the current density is about 0.06 MA/cm${}^{2}$ ($6.7$ $\mu$A in terms of current). 
Such a value is sufficiently small compared with that used in spin-transfer torque switching experiments \cite{yakushiji13}. 
To verify the argument, we perform numerical simulations, where spin-transfer torque, $-H_{\rm s} \mathbf{m} \times (\mathbf{p} \times \mathbf{m})$, is added to the right-hand side of Eq. (\ref{eq:LLG}). 
We fix the values of $H_{\rm K2}=500$ Oe and $H_{\rm K1}=-0.1 H_{\rm K2}=-50$ Oe. 
The unit vector $\mathbf{p}$ along the direction of the magnetization in the reference layer points to the positive $z$ direction. 
Spin polarization $P$ in the spin-transfer torque strength, $H_{\rm s}=\hslash P j/(2eMd)$, is assumed to be $0.5$. 
Figure \ref{fig:fig4}(a) shows time evolution of $\mathbf{m}$ for the current density $j$ of $0.06$ MA/cm${}^{2}$. 
Although the magnetization slightly moves from the initial (stable) state due to spin-transfer torque, the change of the magnetization direction is small compared with that shown in Fig. \ref{fig:fig1}(b). 
Therefore, we do not consider that spin-transfer torque plays a major role in physical reservoir computing, although current cannot be completely zero in experiments. 


For comprehensiveness, however, we also show the magnetization dynamics when the current density $j$ is increased by one order
Figure \ref{fig:fig4}(b) shows the dynamics for $j=0.6$ MA/cm${}^{2}$, where the magnetization switching by spin-transfer torque is observed. 
We note that the current density is sufficiently small compared with that used in typical MTJs in nonvolatile memory \cite{yakushiji13}. 
Nevertheless, the switching is observed because of a small value of the magnetic anisotropy field in the present system. 
We assume that $H_{\rm K2}$ is finite and $|H_{\rm K1}|<H_{\rm K2}$ to make a tilted state of the magnetization [$m_{z}=\pm\sqrt{1-(|H_{\rm K1}|/H_{\rm K2})}$] stable due to the following reason. 
Remind that there are other stable states, such as $m_{z}=\pm 1$ for $H_{\rm K1}>0$ and $m_{z}=0$ for $H_{\rm K1}<0$, when $H_{\rm K2}=0$. 
Note that these states ($m_{z}=\pm 1$ or $m_{z}=0$) are always local extrema of energy landscape. 
Accordingly, once the magnetization saturates to these states, it cannot change the direction even if another input is injected. 
This conclusion can be understood in a different way, where the relaxation time given by Eq. (\ref{eq:relaxation_time}) shows a divergence when $\theta_{\rm i}=0$ ($m_{z}=+1$), $\pi$ ($m_{z}=-1$), or $\pi/2$ ($m_{z}=0$) is substituted. 
On the other hand, for a finite $H_{\rm K2}$, the magnetization can move from the state $m_{z}=\pm\sqrt{1-(|H_{\rm K1}|/H_{\rm K2})}$ 
when an input signal changes the value of $H_{\rm K1}$ and makes the state no longer an extremum. 
We note that the assumption $|H_{\rm K1}|<H_{\rm K2}$ restricts the magnitude of the magnetic field. 
In fact, the magnitude of $\mathbf{H}$ is small due to a small value of $H_{\rm K2}=500$ Oe found in experiments \cite{okada18,sugihara19} and the restriction of $|H_{\rm K1}|<H_{\rm K2}$. 
Since a critical current density destabilizing the magnetization by spin-transfer effect is proportional to the magnitude of the magnetic field, 
a small $\mathbf{H}$ implies that a small current mentioned above might induce a large-amplitude magnetization dynamics. 

In summary, the magnitude of the current density is sufficiently small, and the magnetization dynamics are mainly driven by voltage control of magnetic anisotropy effect. 
The condition to stabilize a tilted state, however, might make the magnitude of the magnetic field, as well as the critical current density of spin-transfer torque switching, small. 
Thus, even a small current may cause nonnegligible dynamics. 
Simultaneously, however, it is practically difficult to increase the current magnitude by one order, and therefore, 
in the present study, we still consider that voltage control of magnetic anisotropy effect is the main driving force of the magnetization dynamics.



\subsection*{Evaluation method of memory capacity}

The memory capacity corresponds to the number of data which can be reproduced from the output data, as mentioned in the main text. 
The evaluation of the memory capacity consists of two processes. 
During the first process called training (or learning), weights are determined to reproduce the target data from the output data. 
In the second process, the reproducibility of the target data defined from other input data is evaluated. 



\begin{figure}
\centerline{\includegraphics[width=1.0\columnwidth]{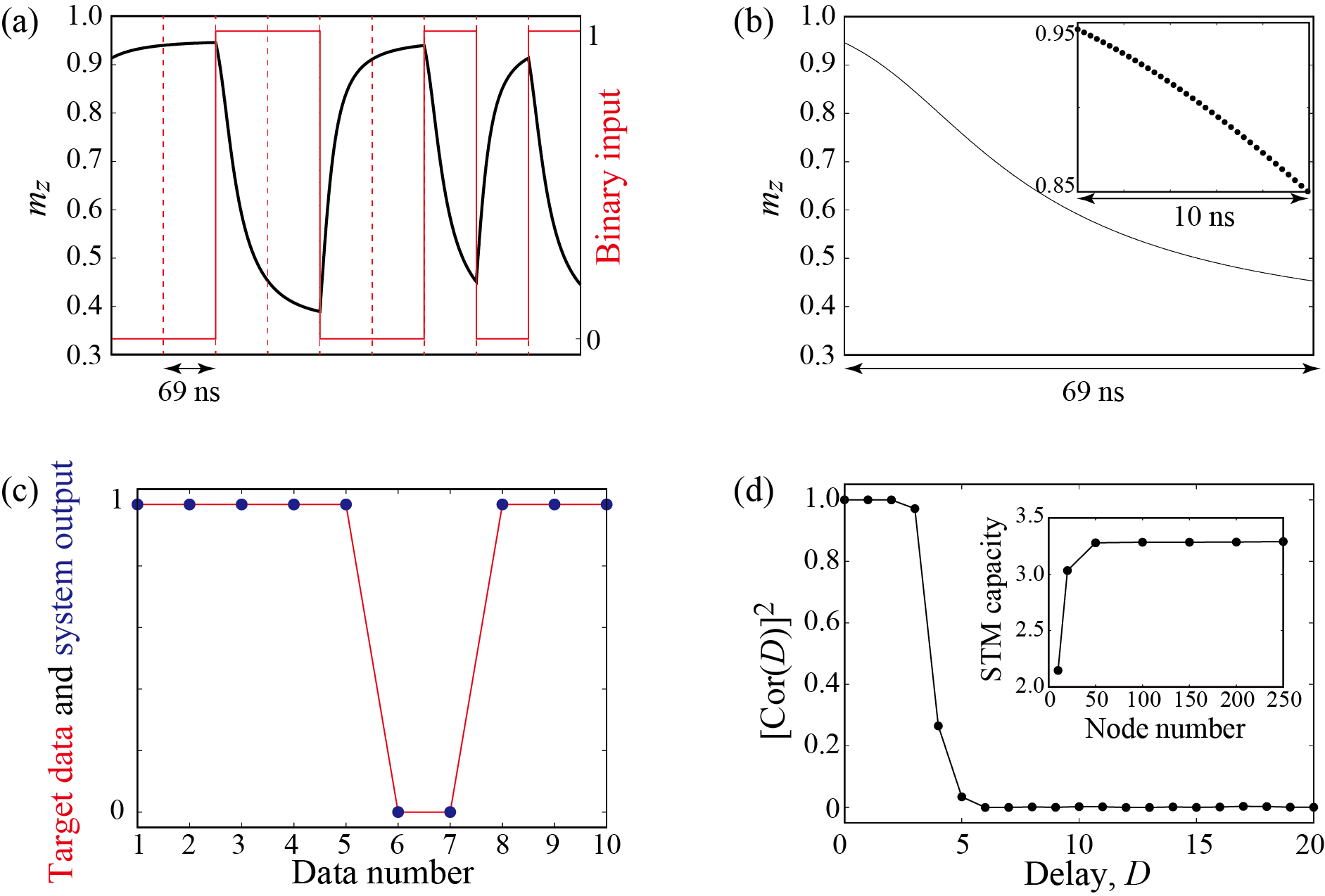}}
\caption{
         (a) An example of the time evolution of $m_{z}$ (black) in the presence of several binary pulses (red). 
             The dotted lines distinguish the input pulse. 
             The pulse width and the first order magnetic anisotropy field are 69 ns and -430 Oe, respectively, 
             where the STM capacity is maximized. 
         (b) An example of $m_{z}$ in the presence of a random input. 
             The dots in the inset shows the definition of the nodes $u_{k,i}$ from $m_{z}$ during a part of an input pulse. 
             The node number is $N_{\rm node}=250$. 
         (c) Examples of the target data $y_{n,D}^{\prime}$ (red line) and the system output $v_{n,D}^{\prime}$ (blue dots) of STM task with $D=1$. 
         (d) Dependence of $[{\rm Cor}(D)]^{2}$ on the delay $D$ for STM task. 
             The node number is $N_{\rm node}=250$. 
             The inset shows the dependence of the STM capacity on the node number. 
         \vspace{-3ex}
         }
\label{fig:fig5}
\end{figure}



Let us first describe the training process. 
We inject the random binary input $b_{k}=0$ or $1$ into MTJ as voltage pulse. 
The number of the random input is $N$. 
The input $b_{k}$ is converted to the first order magnetic anisotropy field through the voltage control of magnetic anisotropy, 
which is described by Eq. (\ref{eq:input_STM_PC}). 
We choose $m_{z}$ as output data, which can be measured experimentally through magnetoresistance effect. 
Figure \ref{fig:fig5}(a) shows an example of the time evolution of $m_{z}$ in the presence of several random binary inputs, 
where the values of the parameters are those at the maximum STM capacity conditions, i.e., 
the pulse width and the first order magnetic anisotropy field are $t_{\rm p}=69$ ns and $H_{\rm K1}^{(1)}=-430$ Oe. 
As can be seen, the injection of the random input drives the dynamics of $m_{z}$. 


The dynamical response $m_{z}(t)$, during the presence of the $k$th input $b_{k}$, is divided into nodes, 
where the number of nodes is $N_{\rm node}$. 
We denote the $i(=1,2,\cdots,N_{\rm node})$th output with respect to the $k$th input as $u_{k,i}=m_{z}(t_{0}+(k-1)t_{\rm p}+i(t_{\rm p}/N_{\rm node}))$, where $t_{0}$ is time for washout. 
The output $u_{k,i}$ is regarded as the status of the $i$th neuron at a discrete time $k$. 
Figure \ref{fig:fig5}(b) shows an example of the time evolution of $m_{z}$ with respect to an input pulse, 
whereas the dots in the inset of the figure are the nodes $u_{k,i}$ defined from $m_{z}$. 
The method to define such virtual neurons is called time-multiplexing method \cite{fujii17,furuta18,tsunegi18}. 
We also introduce bias term $u_{k,N_{\rm node}+1}=1$. 
In the training process, we introduce weight $w_{D,i}$ and evaluate its value to minimize the error, 
\begin{equation}
  \sum_{k=1}^{N}
  \left(
    \sum_{i=1}^{N_{\rm node}+1}
    w_{D,i}
    u_{k,i}
    -
    y_{k,D}
  \right)^{2}, 
  \label{eq:weight_def}
\end{equation}
where, $y_{k,D}$ are the target data defined by Eqs. (\ref{eq:target_STM}) and (\ref{eq:target_PC}). 
For simplicity, we omit the superscripts such as ``STM'' and ``PC'' in the target data 
because the difference in the evaluation method of the STM and PC capacities is merely due to the definition of the target data. 
In the following, we add superscripts or subscripts, such as ``STM'' and ``PC'', when distinguishing quantities related to their capacities are necessary. 
The weight should be introduced for each target data. 
According to the above statement, we denote the weight to evaluate the STM (PC) capacity as $w_{D,i}^{\rm STM(PC)}$, when necessary. 
Also, we note that the weights are different for each delay $D$. 


Once the weights are determined, we inject other random binary inputs $b_{k}^{\prime}$ to the reservoir, 
where the number of the input data is $N^{\prime}$. 
Note that $N^{\prime}$ is not necessarily the same with $N$. 
Here, we use the prime symbol to distinguish the input data from those used in training. 
Similarly, we denote the output and target data with respect to $b_{k}^{\prime}$ as $u_{n,i}^{\prime}$ and $y_{n,D}^{\prime}$, respectively, 
where $n=1,2,\cdots,N^{\prime}$. 
From the output data $u_{n,i}^{\prime}$ and the weight $w_{D,i}$, we define the system output $v_{n,D}^{\prime}$ as 
\begin{equation}
  v_{n,D}^{\prime}
  =
  \sum_{i=1}^{N_{\rm node}+1}
  w_{D,i}
  u_{n,i}^{\prime}. 
  \label{eq:system_output_def}
\end{equation}
Figure \ref{fig:fig5}(c) shows an example of the comparison between the target data $y_{n,D}^{\prime}$ (red line) and the system output $v_{n,D}^{\prime}$ (blue dots) of STM task with $D=1$. 
It is shown that the system output well reproduces the target data. 
The reproducibility of the target data is quantified from the correlation coefficient ${\rm Cor}(D)$ between $y_{n,D}^{\prime}$ and $v_{n,D}^{\prime}$ defined as 
\begin{equation}
  {\rm Cor}(D)
  \equiv
  \frac{ \sum_{n=1}^{N^{\prime}} \left( y_{n,D}^{\prime} - \langle y_{n,D}^{\prime} \rangle \right) \left( v_{n,D}^{\prime} - \langle v_{n,D}^{\prime} \rangle \right) }
    { \sqrt{ \sum_{n=1}^{N^{\prime}} \left( y_{n,D}^{\prime} - \langle y_{n,D}^{\prime} \rangle \right)^{2} \sum_{n=1}^{N^{\prime}} \left( v_{n,D}^{\prime} - \langle v_{n,D}^{\prime} \rangle \right)^{2} } }, 
\end{equation}
where $\langle \cdots \rangle$ represents the averaged value. 
Note that the correlation coefficients are defined for each delay $D$. 
We also note that the correlation coefficients are defined for each kind of capacity, as in the case of the weights and target data. 
In general, $[{\rm Cor}(D)]^{2} \le 1$, where $[{\rm Cor}(D)]^{2}=1$ holds only when the system output completely reproduces the target data. 
Figure \ref{fig:fig5}(d) shows an example of the dependence of $[{\rm Cor}(D)]^{2}$ for STM task on the delay $D$. 
The results implies that the reservoir well reproduces the target data until $D=3$, whereas the reproducibility drastically decreases with the delay $D$ increasing. 
The STM and PC capacities, $C_{\rm STM}$ and $C_{\rm PC}$, are defined as 
\begin{equation}
  C
  =
  \sum_{D=1}^{D_{\rm max}}
  \left[
    {\rm Cor}(D)
  \right]^{2}.
  \label{eq:memory_capacity_def}
\end{equation}
Note that the definition of the memory capacity obeys, for example, Refs. \cite{goto21,furuta18,tsunegi18,tsunegi19}, 
where the memory capacity in Eq. (\ref{eq:memory_capacity_def}) is defined by the correlation coefficients starting from $D=1$. 
In some papers such as Refs. \cite{fujii17,akashi20}, however, the square of the correlation coefficient at $D=0$ is added to the right-hand side of Eq. (\ref{eq:memory_capacity_def}). 


In the present study, we introduce $N_{\rm node}=250$ nodes and use $N=1000$ and $N^{\prime}=1000$ random binary pulses for training of the weight and evaluation of the memory capacity, respectively. 
The number of nodes is chosen so that the value of the capacity saturates with the number of nodes increasing; see the inset of Fig. \ref{fig:fig5}(d). 
We also use 300 random binary pulses before the training and between training and evaluation for washout. 
The maximum delay $D_{\rm max}$ is $20$. 
Note that the value of each node should be sampled within a few hundred picosecond: 
specifically, in the case of an example shown in Fig. \ref{fig:fig2}(c), it is necessary to sample data within $t_{\rm p}/N_{\rm node}=69 {\rm ns}/250 \simeq 276$ ps. 
We emphasize that it is experimentally possible to sample data within such a short time. 
For example, in Ref. \cite{tsunegi18}, $t_{\rm p}=20$ ns and $N_{\rm node}=200$ were used, where the sampling step is $100$ ps. 


\subsection*{NARMA task}

The evaluation procedure of the NMSE in NARMA task is similar to that of the memory capacity. 
The binary input data, $b_{k}=0$ or $1$, in the evaluation of the memory capacity are replaced by uniform random number $r_{k}$ in $(0,1)$. 
The variable $z_{k}$ in Eq. (\ref{eq:NARMA2_def}) is generally defined as $z_{k}=\mu+\sigma r_{k}$ \cite{akashi20}, 
where the parameters $\mu$ and $\sigma$ are determined to make $z_{k}$ be in $(0,0.2)$ \cite{fujii17}. 
As in the case of the evaluation of the memory capacity, the evaluation of the NMSE consists of two procedures. 
The first procedure is the training, where the weight is determined to reproduce the target data from the output data $u_{k,i}$. 
Secondly, we evaluate the reproducibility of another set of the target data from the system output $v_{n}^{\rm NARMA2}$ 
defined from the weight and the output data. 
Then, the NMSE can be evaluated. 
Note that some papers \cite{nakayama16,nomura19,akashi20} define the NMSE in a slightly different way, where $\sum_{n=1}^{N^{\prime}}\left( y_{n}^{\rm NARMA2} \right)^{2}$ in the denominator of Eq. (\ref{eq:NMSE_def}) 
is replaced by $\sum_{n=1}^{N^{\prime}} \left( y_{n}^{\rm NARMA2} - \overline{y}^{\rm NARMA2} \right)^{2}$, 
where $\overline{y}^{\rm NARMA2}$ is the average of the target data $y_{n}^{\rm NARMA2}$. 
In this work, we use the definition given by Eq. (\ref{eq:NMSE_def}), which is used, for example, in Refs. \cite{nakajima15,fujii17,goto21}. 


\subsection*{Evaluation of Lyapunov exponent}

We evaluated the conditional Lyapunov exponent as follows \cite{taniguchi20AIP}. 
The LLG equation was solved by the fourth-order Runge-Kutta method with time increment of $\Delta t=1$ ps. 
We added perturbations $\delta\theta$ and $\delta\phi$ with $\epsilon=\sqrt{\delta\theta^{2}+\delta\varphi^{2}}=10^{-5}$ to $\theta(t)$ and $\varphi(t)$ at time $t$. 
Let us denote the perturbed $\theta(t)$ and $\varphi(t)$ as $\theta^{\prime}(t)$ and $\varphi^{\prime}(t)$, respectively. 
Solving the LLG equation from time $t$ to $t+\Delta t$, 
the time evolution of the perturbation is obtained as 
$\epsilon^{\prime}(t)=\sqrt{[\theta^{\prime}(t+\Delta t)-\theta(t+\Delta t)]^{2}+[\varphi^{\prime}(t+\Delta t)-\varphi(t+\Delta t)]^{2}}$. 
A temporal Lyapunov exponent is obtained as $\lambda(t)=(1/\Delta t)\log [\epsilon^{\prime}(t)/\epsilon]$. 
Repeating the procedure, the temporal Lyapunov exponent is averaged as 
$\lambda(\mathcal{N})=(1/\mathcal{N})\sum_{i=1}^{\mathcal{N}}\lambda(t_{i})=[1/(\mathcal{N} \Delta t)]\sum_{i=1}^{\mathcal{N}}\log\{\epsilon^{\prime}[t_{0}+(i-1) \Delta t]/\epsilon\}$, 
where $t_{0}$ is time at which the first random input is injected, whereas $\mathcal{N}$ is the number of averaging. 
The Lyapunov exponent is given by $\lambda=\lim_{\mathcal{N} \to \infty}\lambda(\mathcal{N})$. 
In the present study, we used the time range same as that used in the evaluations of the memory capacity and the NMSE and added uniform random input. 
Hence, notice that $\mathcal{N}=\mathcal{M}t_{\rm p}/\Delta t$ depends on the pulse width, where $\mathcal{M}$ is the total number of the random inputs including washout, training, and evaluation. 
We confirmed that $\lambda(\mathcal{N})$ monotonically saturates to zero; 
at least, $|\lambda(\mathcal{N})|$ is one or two orders of magnitudes smaller than $1/t_{\rm p}$. 
Thus, the expansion rate of the perturbation, $1/\lambda(\mathcal{N})$, is much slower than the injection rate of the input signal. 
Considering these facts, we concluded that the largest Lyapunov exponent can be regarded as zero, and therefore, chaos is absent. 
Note that the absence of chaos in the present system relates to the facts that the free layer is axially symmetric and the applied voltage modifies the perpendicular anisotropy only. 
When there are factors breaking the symmetry, such as spin-transfer torque with an in-plane spin polarization, chaos will appear \cite{akashi20}. 




\begin{thebibliography}{10}
\urlstyle{rm}
\expandafter\ifx\csname url\endcsname\relax
  \def\url#1{\texttt{#1}}\fi
\expandafter\ifx\csname urlprefix\endcsname\relax\def\urlprefix{URL }\fi
\expandafter\ifx\csname doiprefix\endcsname\relax\def\doiprefix{DOI: }\fi
\providecommand{\bibinfo}[2]{#2}
\providecommand{\eprint}[2][]{\url{#2}}

\bibitem{torrejon17}
\bibinfo{author}{Torrejon, J.} \emph{et~al.}
\newblock \bibinfo{journal}{\bibinfo{title}{Neuromorphic computing with
  nanoscale spintronic oscillators}}.
\newblock {\emph{\JournalTitle{Nature}}} \textbf{\bibinfo{volume}{547}},
  \bibinfo{pages}{428} (\bibinfo{year}{2017}).

\bibitem{borders17}
\bibinfo{author}{Borders, W.~A.} \emph{et~al.}
\newblock \bibinfo{journal}{\bibinfo{title}{Analogue spin-orbit torque device
  for artificial-neural-network-based associative memory operation}}.
\newblock {\emph{\JournalTitle{Appl. Phys. Express}}}
  \textbf{\bibinfo{volume}{10}}, \bibinfo{pages}{013007}
  (\bibinfo{year}{2017}).

\bibitem{kudo17}
\bibinfo{author}{Kudo, K.} \& \bibinfo{author}{Morie, T.}
\newblock \bibinfo{journal}{\bibinfo{title}{Self-feedback electrically coupled
  spin-{Hall} oscillator array for pattern-matching operation}}.
\newblock {\emph{\JournalTitle{Appl. Phys. Express}}}
  \textbf{\bibinfo{volume}{10}}, \bibinfo{pages}{043001}
  (\bibinfo{year}{2017}).

\bibitem{grollier20}
\bibinfo{author}{Grollier, J.} \emph{et~al.}
\newblock \bibinfo{journal}{\bibinfo{title}{Neuromorphic spintronics}}.
\newblock {\emph{\JournalTitle{Nat. Electron.}}} \textbf{\bibinfo{volume}{3}},
  \bibinfo{pages}{360} (\bibinfo{year}{2020}).

\bibitem{maas02}
\bibinfo{author}{Maas, W.}, \bibinfo{author}{Natschl\"ager, T.} \&
  \bibinfo{author}{Markram, H.}
\newblock \bibinfo{journal}{\bibinfo{title}{Real-{Time} {Computing} {Without}
  {Stable} {States}: {A} {New} {Framework} for {Neural} {Computation} {Based}
  on {Perturbations}}}.
\newblock {\emph{\JournalTitle{Neural Comput.}}} \textbf{\bibinfo{volume}{14}},
  \bibinfo{pages}{2531} (\bibinfo{year}{2002}).

\bibitem{jaeger04}
\bibinfo{author}{Jaeger, H.} \& \bibinfo{author}{Haas, H.}
\newblock \bibinfo{journal}{\bibinfo{title}{Harnessing {Nonlinearity}:
  {Predicting} {Chaotic} {Systems} and {Saving} {Energy} in {Wireless}
  {Communication}}}.
\newblock {\emph{\JournalTitle{Science}}} \textbf{\bibinfo{volume}{304}},
  \bibinfo{pages}{78} (\bibinfo{year}{2004}).

\bibitem{verstraeten07}
\bibinfo{author}{Verstraeten, D.}, \bibinfo{author}{Schrauwen, B.},
  \bibinfo{author}{D'Haene, M.} \& \bibinfo{author}{Stroobandt, D.}
\newblock \bibinfo{journal}{\bibinfo{title}{An experimental unification of
  reservoir computing methods}}.
\newblock {\emph{\JournalTitle{Neural Netw.}}} \textbf{\bibinfo{volume}{20}},
  \bibinfo{pages}{391} (\bibinfo{year}{2007}).

\bibitem{hermans10}
\bibinfo{author}{Hermans, M.} \& \bibinfo{author}{Schrauwen, B.}
\newblock \bibinfo{journal}{\bibinfo{title}{Memory in linear recurrent neural
  networks in continuous time}}.
\newblock {\emph{\JournalTitle{Neural Netw.}}} \textbf{\bibinfo{volume}{23}},
  \bibinfo{pages}{341} (\bibinfo{year}{2010}).

\bibitem{appeltant11}
\bibinfo{author}{Appeltant, L.} \emph{et~al.}
\newblock \bibinfo{journal}{\bibinfo{title}{Information processing using a
  single dynamical node as complex system}}.
\newblock {\emph{\JournalTitle{Nat. Commun.}}} \textbf{\bibinfo{volume}{2}},
  \bibinfo{pages}{468} (\bibinfo{year}{2011}).

\bibitem{paquot12}
\bibinfo{author}{Paquot, Y.} \emph{et~al.}
\newblock \bibinfo{journal}{\bibinfo{title}{Optoelectronic {Reservoir}
  {Computing}}}.
\newblock {\emph{\JournalTitle{Sci. Rep.}}} \textbf{\bibinfo{volume}{2}},
  \bibinfo{pages}{287} (\bibinfo{year}{2012}).

\bibitem{brunner13}
\bibinfo{author}{Brunner, D.}, \bibinfo{author}{Soriano, M.~C.},
  \bibinfo{author}{Mirasso, C.~R.} \& \bibinfo{author}{Fischer, I.}
\newblock \bibinfo{journal}{\bibinfo{title}{Parallel photonic information
  processing at gigabyte per second data rates using trasient states}}.
\newblock {\emph{\JournalTitle{Nat. Commun.}}} \textbf{\bibinfo{volume}{4}},
  \bibinfo{pages}{1364} (\bibinfo{year}{2013}).

\bibitem{nakajima15}
\bibinfo{author}{Nakajima, K.}, \bibinfo{author}{Hauser, H.},
  \bibinfo{author}{Li, T.} \& \bibinfo{author}{Pfeifer, R.}
\newblock \bibinfo{journal}{\bibinfo{title}{Information processing via physical
  soft body}}.
\newblock {\emph{\JournalTitle{Sci. Rep.}}} \textbf{\bibinfo{volume}{5}},
  \bibinfo{pages}{10487} (\bibinfo{year}{2015}).

\bibitem{nakayama16}
\bibinfo{author}{Nakayama, J.}, \bibinfo{author}{Kanno, K.} \&
  \bibinfo{author}{Uchida, A.}
\newblock \bibinfo{journal}{\bibinfo{title}{Laser dynamical reservoir computing
  with consistency: an approach of a chaos mask signal}}.
\newblock {\emph{\JournalTitle{Opt. Express}}} \textbf{\bibinfo{volume}{24}},
  \bibinfo{pages}{8679--8692} (\bibinfo{year}{2016}).

\bibitem{sande17}
\bibinfo{author}{der Sande, G.~V.}, \bibinfo{author}{Brunner, D.} \&
  \bibinfo{author}{Soriano, M.~C.}
\newblock \bibinfo{journal}{\bibinfo{title}{Advances in photonic reservoir
  computing}}.
\newblock {\emph{\JournalTitle{Nanophotonics}}} \textbf{\bibinfo{volume}{6}},
  \bibinfo{pages}{561--576} (\bibinfo{year}{2017}).

\bibitem{fujii17}
\bibinfo{author}{Fujii, K.} \& \bibinfo{author}{Nakajima, K.}
\newblock \bibinfo{journal}{\bibinfo{title}{Harnessing {Disordered}-{Ensemble}
  {Quantum} {Dynamics} for {Machine} {Learning}}}.
\newblock {\emph{\JournalTitle{Phys. Rev. Applied}}}
  \textbf{\bibinfo{volume}{8}}, \bibinfo{pages}{024030} (\bibinfo{year}{2017}).

\bibitem{dion18}
\bibinfo{author}{Dion, G.}, \bibinfo{author}{Mejaouri, S.} \&
  \bibinfo{author}{Sylvestre, J.}
\newblock \bibinfo{journal}{\bibinfo{title}{Reservoir computing with a single
  delay-coupled non-linear mechanical oscillator}}.
\newblock {\emph{\JournalTitle{J. Appl. Phys.}}}
  \textbf{\bibinfo{volume}{124}}, \bibinfo{pages}{152132}
  (\bibinfo{year}{2018}).

\bibitem{nakajima20}
\bibinfo{author}{Nakajima, K.}
\newblock \bibinfo{journal}{\bibinfo{title}{Physical reservoir computing - an
  introductory perspective}}.
\newblock {\emph{\JournalTitle{Jpn. J. Appl. Phys.}}}
  \textbf{\bibinfo{volume}{59}}, \bibinfo{pages}{060501}
  (\bibinfo{year}{2020}).

\bibitem{goto21}
\bibinfo{author}{Goto, K.}, \bibinfo{author}{Nakajima, K.} \&
  \bibinfo{author}{Notsu, H.}
\newblock \bibinfo{journal}{\bibinfo{title}{Twin vortex computer in fluid
  flow}}.
\newblock {\emph{\JournalTitle{New J. Phys.}}} \textbf{\bibinfo{volume}{23}},
  \bibinfo{pages}{063051} (\bibinfo{year}{202}).

\bibitem{nakajima21}
\bibinfo{editor}{Nakajima, K.} \& \bibinfo{editor}{Fischer, I.} (eds.)
  \emph{\bibinfo{title}{Reservoir Computing: Theory, Physical Implementations,
  and Applications}} (\bibinfo{publisher}{Springer, Singapore},
  \bibinfo{year}{2021}).

\bibitem{furuta18}
\bibinfo{author}{Furuta, T.} \emph{et~al.}
\newblock \bibinfo{journal}{\bibinfo{title}{Macromagnetic {Simulation} for
  {Reservoir} {Computing} {Utilizing} {Spin} {Dynamics} in {Magnetic} {Tunnel}
  {Junctions}}}.
\newblock {\emph{\JournalTitle{Phys. Rev. Applied}}}
  \textbf{\bibinfo{volume}{10}}, \bibinfo{pages}{034063}
  (\bibinfo{year}{2018}).

\bibitem{tsunegi18}
\bibinfo{author}{Tsunegi, S.} \emph{et~al.}
\newblock \bibinfo{journal}{\bibinfo{title}{Evaluation of memory capacity of
  spin torque oscillator for recurrent neural networks}}.
\newblock {\emph{\JournalTitle{Jpn. J. Appl. Phys.}}}
  \textbf{\bibinfo{volume}{57}}, \bibinfo{pages}{120307}
  (\bibinfo{year}{2018}).

\bibitem{bourianoff18}
\bibinfo{author}{Bourianoff, G.}, \bibinfo{author}{Pinna, D.},
  \bibinfo{author}{Sitte, M.} \& \bibinfo{author}{Everschor-Sitte, K.}
\newblock \bibinfo{journal}{\bibinfo{title}{Potential implementation of
  reservoir computing models based on magnetic skyrmions}}.
\newblock {\emph{\JournalTitle{AIP Adv.}}} \textbf{\bibinfo{volume}{8}},
  \bibinfo{pages}{055602} (\bibinfo{year}{2018}).

\bibitem{nakane18}
\bibinfo{author}{Nakane, R.}, \bibinfo{author}{Tanaka, G.} \&
  \bibinfo{author}{Hirose, A.}
\newblock \bibinfo{journal}{\bibinfo{title}{Reservoir {Computing} {With} {Spin}
  {Waves} {Excited} in a {Garnet} {Film}}}.
\newblock {\emph{\JournalTitle{IEEE Access}}} \textbf{\bibinfo{volume}{6}},
  \bibinfo{pages}{4462} (\bibinfo{year}{2018}).

\bibitem{markovic19}
\bibinfo{author}{Markovi\'c, D.} \emph{et~al.}
\newblock \bibinfo{journal}{\bibinfo{title}{Reservoir computing with the
  frequency, phase, and amplitude of spin-torque nano-oscillators}}.
\newblock {\emph{\JournalTitle{Appl. Phys. Lett.}}}
  \textbf{\bibinfo{volume}{114}}, \bibinfo{pages}{012409}
  (\bibinfo{year}{2019}).

\bibitem{tsunegi19}
\bibinfo{author}{Tsunegi, S.} \emph{et~al.}
\newblock \bibinfo{journal}{\bibinfo{title}{Physical reservoir computing based
  on spin torque oscillator with forced synchronization}}.
\newblock {\emph{\JournalTitle{Appl. Phys. Lett.}}}
  \textbf{\bibinfo{volume}{114}}, \bibinfo{pages}{164101}
  (\bibinfo{year}{2019}).

\bibitem{riou19}
\bibinfo{author}{Riou, M.} \emph{et~al.}
\newblock \bibinfo{journal}{\bibinfo{title}{Temporal {Patter} {Recognition}
  with {Delayed}-{Feedback} {Spin}-{Torque} {Nano}-{Oscillators}}}.
\newblock {\emph{\JournalTitle{Phys. Rev. Applied}}}
  \textbf{\bibinfo{volume}{12}}, \bibinfo{pages}{024049}
  (\bibinfo{year}{2019}).

\bibitem{nomura19}
\bibinfo{author}{Nomura, H.} \emph{et~al.}
\newblock \bibinfo{journal}{\bibinfo{title}{Reservoir computing with
  dipole-coupled nanomagnets}}.
\newblock {\emph{\JournalTitle{Jpn. J. Appl. Phys.}}}
  \textbf{\bibinfo{volume}{58}}, \bibinfo{pages}{070901}
  (\bibinfo{year}{2019}).

\bibitem{yamaguchi20}
\bibinfo{author}{Yamaguchi, T.} \emph{et~al.}
\newblock \bibinfo{journal}{\bibinfo{title}{Periodic structure of memory
  function in spintronics reservoir with feedback current}}.
\newblock {\emph{\JournalTitle{Phys. Rev. Research}}}
  \textbf{\bibinfo{volume}{2}}, \bibinfo{pages}{023389} (\bibinfo{year}{2020}).

\bibitem{yamaguchi20srep}
\bibinfo{author}{Yamaguchi, T.} \emph{et~al.}
\newblock \bibinfo{journal}{\bibinfo{title}{Step-like dependence of memory
  function on pulse width in spintronics reservoir computing}}.
\newblock {\emph{\JournalTitle{Sci. Rep.}}} \textbf{\bibinfo{volume}{10}},
  \bibinfo{pages}{19536} (\bibinfo{year}{2020}).

\bibitem{akashi20}
\bibinfo{author}{Akashi, N.} \emph{et~al.}
\newblock \bibinfo{journal}{\bibinfo{title}{Input-driven bifurcations and
  information processing capacity in spintronics reservoirs}}.
\newblock {\emph{\JournalTitle{Phys. Rev. Research}}}
  \textbf{\bibinfo{volume}{2}}, \bibinfo{pages}{043303} (\bibinfo{year}{2020}).

\bibitem{slonczewski96}
\bibinfo{author}{Slonczewski, J.~C.}
\newblock \bibinfo{journal}{\bibinfo{title}{Current-driven excitation of
  magnetic multilayers}}.
\newblock {\emph{\JournalTitle{J. Magn. Magn. Mater.}}}
  \textbf{\bibinfo{volume}{159}}, \bibinfo{pages}{L1} (\bibinfo{year}{1996}).

\bibitem{berger96}
\bibinfo{author}{Berger, L.}
\newblock \bibinfo{journal}{\bibinfo{title}{Emission of spin waves by a
  magnetic multilayer traversed by a current}}.
\newblock {\emph{\JournalTitle{Phys. Rev. B}}} \textbf{\bibinfo{volume}{54}},
  \bibinfo{pages}{9353} (\bibinfo{year}{1996}).

\bibitem{weisheit07}
\bibinfo{author}{Weisheit, M.} \emph{et~al.}
\newblock \bibinfo{journal}{\bibinfo{title}{Electric {Field}-{Induced}
  {Modification} of {Magnetism} in {Thin}-{Film} {Ferromagnets}}}.
\newblock {\emph{\JournalTitle{Science}}} \textbf{\bibinfo{volume}{315}},
  \bibinfo{pages}{349} (\bibinfo{year}{2007}).

\bibitem{duan08}
\bibinfo{author}{Duan, C.-G.} \emph{et~al.}
\newblock \bibinfo{journal}{\bibinfo{title}{Surface {Magnetoelectric} {Effect}
  in {Ferromagnetic} {Metal} {Films}}}.
\newblock {\emph{\JournalTitle{Phys. Rev. Lett.}}}
  \textbf{\bibinfo{volume}{101}}, \bibinfo{pages}{137201}
  (\bibinfo{year}{2008}).

\bibitem{maruyama09}
\bibinfo{author}{Maruyama, T.} \emph{et~al.}
\newblock \bibinfo{journal}{\bibinfo{title}{Large voltage-induced magnetic
  anisotropy change in a few atomic layers of iron}}.
\newblock {\emph{\JournalTitle{Nat. Nanotechnol.}}}
  \textbf{\bibinfo{volume}{4}}, \bibinfo{pages}{158} (\bibinfo{year}{2009}).

\bibitem{nakamura09}
\bibinfo{author}{Nakamura, K.} \emph{et~al.}
\newblock \bibinfo{journal}{\bibinfo{title}{Giant {Modification} of the
  {Magnetocrystalline} {Anisotropy} in {Transition}-{Metal} {Monolayers} by an
  {External} {Electric} {Field}}}.
\newblock {\emph{\JournalTitle{Phys. Rev. Lett.}}}
  \textbf{\bibinfo{volume}{102}}, \bibinfo{pages}{187201}
  (\bibinfo{year}{2009}).

\bibitem{tsujikawa09}
\bibinfo{author}{Tsujikawa, M.} \& \bibinfo{author}{Oda, T.}
\newblock \bibinfo{journal}{\bibinfo{title}{Finite {Electric} {Field} {Effects}
  in the {Large} {Perpendicular} {Magnetic} {Anisotropy} {Surface}
  {Pt}/{Fe}/{Pt}(001): {A} {First}-{Principles} {Study}}}.
\newblock {\emph{\JournalTitle{Phys. Rev. Lett.}}}
  \textbf{\bibinfo{volume}{102}}, \bibinfo{pages}{247203}
  (\bibinfo{year}{2009}).

\bibitem{shiota09}
\bibinfo{author}{Shiota, Y.} \emph{et~al.}
\newblock \bibinfo{journal}{\bibinfo{title}{Voltage-{Assisted} {Magnetization}
  {Switching} in {Ultrahin} {Fe}${}_{80}$ {Co}${}_{20}$ {Alloy} {Layers}}}.
\newblock {\emph{\JournalTitle{Appl. Phys. Express}}}
  \textbf{\bibinfo{volume}{2}}, \bibinfo{pages}{063001} (\bibinfo{year}{2009}).

\bibitem{nozaki10}
\bibinfo{author}{Nozaki, T.}, \bibinfo{author}{Shiota, Y.},
  \bibinfo{author}{Shiraishi, M.}, \bibinfo{author}{Shinjo, T.} \&
  \bibinfo{author}{Suzuki, Y.}
\newblock \bibinfo{journal}{\bibinfo{title}{Voltage-induced perpendicular
  magnetic anisotropy change in magnetic tunnel junctions}}.
\newblock {\emph{\JournalTitle{Appl. Phys. Lett.}}}
  \textbf{\bibinfo{volume}{96}}, \bibinfo{pages}{022506}
  (\bibinfo{year}{2010}).

\bibitem{shiota11}
\bibinfo{author}{Shiota, Y.} \emph{et~al.}
\newblock \bibinfo{journal}{\bibinfo{title}{Induction of coherent magnetization
  switching in a few atomic layers of {Fe}{Co} using voltage pulses}}.
\newblock {\emph{\JournalTitle{Nat. Mater.}}} \textbf{\bibinfo{volume}{11}},
  \bibinfo{pages}{39} (\bibinfo{year}{2011}).

\bibitem{wang11}
\bibinfo{author}{Wang, W.-G.}, \bibinfo{author}{Li, M.},
  \bibinfo{author}{Hageman, S.} \& \bibinfo{author}{Chien, C.~L.}
\newblock \bibinfo{journal}{\bibinfo{title}{Electric-field-assisted switching
  in magnetic tunnel junctions}}.
\newblock {\emph{\JournalTitle{Nat. Mater.}}} \textbf{\bibinfo{volume}{11}},
  \bibinfo{pages}{64} (\bibinfo{year}{2011}).

\bibitem{shiota12}
\bibinfo{author}{Shiota, Y.} \emph{et~al.}
\newblock \bibinfo{journal}{\bibinfo{title}{Pulse voltage-induced dynamic
  magnetization switching in magnetic tunnel junctions with high
  resistance-area product}}.
\newblock {\emph{\JournalTitle{Appl. Phys. Lett.}}}
  \textbf{\bibinfo{volume}{101}}, \bibinfo{pages}{102406}
  (\bibinfo{year}{2012}).

\bibitem{kanai12}
\bibinfo{author}{Kanai, S.} \emph{et~al.}
\newblock \bibinfo{journal}{\bibinfo{title}{Electric field-induced
  magnetization reversal in a perpendicular-anisotropy {Co}{Fe}{B}-{Mg}{O}
  magnetic tunnel junction}}.
\newblock {\emph{\JournalTitle{Appl. Phys. Lett.}}}
  \textbf{\bibinfo{volume}{101}}, \bibinfo{pages}{122403}
  (\bibinfo{year}{2012}).

\bibitem{grezes16}
\bibinfo{author}{Grezes, C.} \emph{et~al.}
\newblock \bibinfo{journal}{\bibinfo{title}{Ultra-low switching energy and
  scaling in electric-field-controlled nanoscale magnetic tunnel junctions with
  high resistance-area product}}.
\newblock {\emph{\JournalTitle{Appl. Phys. Lett.}}}
  \textbf{\bibinfo{volume}{108}}, \bibinfo{pages}{012403}
  (\bibinfo{year}{2016}).

\bibitem{nozaki17}
\bibinfo{author}{Nozaki, T.} \emph{et~al.}
\newblock \bibinfo{journal}{\bibinfo{title}{Highly effcient voltage control of
  spin and enhanced interfacial perpendicular magnetic anisotropy in
  iridium-doped {Fe}/{Mg}{O} magnetic tunnel junctions}}.
\newblock {\emph{\JournalTitle{NPG Asia Mater.}}} \textbf{\bibinfo{volume}{9}},
  \bibinfo{pages}{e451} (\bibinfo{year}{2017}).

\bibitem{miwa17}
\bibinfo{author}{Miwa, S.} \emph{et~al.}
\newblock \bibinfo{journal}{\bibinfo{title}{Voltage controlled interfacial
  magnetism through platinum orbits}}.
\newblock {\emph{\JournalTitle{Nat. Commun.}}} \textbf{\bibinfo{volume}{8}},
  \bibinfo{pages}{15848} (\bibinfo{year}{2017}).

\bibitem{okada18}
\bibinfo{author}{Okada, A.}, \bibinfo{author}{Kanai, S.},
  \bibinfo{author}{Fukami, S.}, \bibinfo{author}{Sato, H.} \&
  \bibinfo{author}{Ohno, H.}
\newblock \bibinfo{journal}{\bibinfo{title}{Electric-field effects on the easy
  cone angle of the easy-cone state in {Co}{Fe}{B}/{Mg}{O} investigated by
  ferromagnetic resonance}}.
\newblock {\emph{\JournalTitle{Appl. Phys. Lett.}}}
  \textbf{\bibinfo{volume}{112}}, \bibinfo{pages}{172402}
  (\bibinfo{year}{2018}).

\bibitem{sugihara19}
\bibinfo{author}{Sugihara, A.} \emph{et~al.}
\newblock \bibinfo{journal}{\bibinfo{title}{Evaluation of higher order magnetic
  anisotropy in a perpendicularly magnetized epitaxial ultrathin {Fe} layer and
  its applied voltage dependence}}.
\newblock {\emph{\JournalTitle{Jpn. J. Appl. Phys.}}}
  \textbf{\bibinfo{volume}{58}}, \bibinfo{pages}{090905}
  (\bibinfo{year}{2019}).

\bibitem{yamamoto19}
\bibinfo{author}{Yamamoto, T.} \emph{et~al.}
\newblock \bibinfo{journal}{\bibinfo{title}{Improvement of write error rate in
  voltage-driven magnetization switching}}.
\newblock {\emph{\JournalTitle{J. Phys. D: Appl. Phys.}}}
  \textbf{\bibinfo{volume}{52}}, \bibinfo{pages}{164001}
  (\bibinfo{year}{2019}).

\bibitem{nozaki20}
\bibinfo{author}{Nozaki, T.} \emph{et~al.}
\newblock \bibinfo{journal}{\bibinfo{title}{Voltage-cotrolled magnetic
  anisotropy in an ultrathin {Ir}-doped {Fe} layer with a {Co}{Fe} termination
  layer}}.
\newblock {\emph{\JournalTitle{APL Mater.}}} \textbf{\bibinfo{volume}{8}},
  \bibinfo{pages}{011108} (\bibinfo{year}{2020}).

\bibitem{yakata09}
\bibinfo{author}{Yakata, S.} \emph{et~al.}
\newblock \bibinfo{journal}{\bibinfo{title}{Influnence of perpendicular
  magnetic anisotropy on spin-transfer switching current in
  {C}o{F}e{B}/{M}g{O}/{C}o{F}e{B} magnetic tunnel junctions}}.
\newblock {\emph{\JournalTitle{J. Appl. Phys.}}}
  \textbf{\bibinfo{volume}{105}}, \bibinfo{pages}{07D131}
  (\bibinfo{year}{2009}).

\bibitem{ikeda10}
\bibinfo{author}{Ikeda, S.} \emph{et~al.}
\newblock \bibinfo{journal}{\bibinfo{title}{A perpendicular-anisotropy
  {C}o{F}e{B}-{M}g{O} magnetic tunnel junction}}.
\newblock {\emph{\JournalTitle{Nat. Mater.}}} \textbf{\bibinfo{volume}{9}},
  \bibinfo{pages}{721} (\bibinfo{year}{2010}).

\bibitem{kubota12}
\bibinfo{author}{Kubota, H.} \emph{et~al.}
\newblock \bibinfo{journal}{\bibinfo{title}{Enhancement of perpendicular
  magnetic anisotropy in {F}e{B} free layers using a thin {M}g{O} cap layer}}.
\newblock {\emph{\JournalTitle{J. Appl. Phys.}}}
  \textbf{\bibinfo{volume}{111}}, \bibinfo{pages}{07C723}
  (\bibinfo{year}{2012}).

\bibitem{yakushiji13}
\bibinfo{author}{Yakushiji, K.}, \bibinfo{author}{Fukushima, A.},
  \bibinfo{author}{Kubota, H.}, \bibinfo{author}{Konoto, M.} \&
  \bibinfo{author}{Yuasa, S.}
\newblock \bibinfo{journal}{\bibinfo{title}{Ultralow-{Voltage}
  {Spin}-{Transfer} {Switching} in {Perpendicularly} {Magnetized} {Magnetic}
  {Tunnel} {Junctions} with {Synthetic} {Antiferromagnetic} {Reference}
  {Layer}}}.
\newblock {\emph{\JournalTitle{Appl. Phys. Express}}}
  \textbf{\bibinfo{volume}{6}}, \bibinfo{pages}{113006} (\bibinfo{year}{2013}).

\bibitem{matsumoto15}
\bibinfo{author}{Matsumoto, R.}, \bibinfo{author}{Arai, H.},
  \bibinfo{author}{Yuasa, S.} \& \bibinfo{author}{Imamura, H.}
\newblock \bibinfo{journal}{\bibinfo{title}{Spin-transfer-torque switching in a
  spin-valve nanopillar with a conically magnetized free layer}}.
\newblock {\emph{\JournalTitle{Appl. Phys. Express}}}
  \textbf{\bibinfo{volume}{8}}, \bibinfo{pages}{063007} (\bibinfo{year}{2015}).

\bibitem{atiya00}
\bibinfo{author}{Atiya, A.~F.}
\newblock \bibinfo{journal}{\bibinfo{title}{New {Results} on {Recurrent}
  {Network} {Training}: {Unifying} the {Algorithms} and {Accelerating}
  {Convergence}}}.
\newblock {\emph{\JournalTitle{IEEE Trans. Neural. Netw.}}}
  \textbf{\bibinfo{volume}{11}}, \bibinfo{pages}{697} (\bibinfo{year}{2000}).

\bibitem{bertschinger04}
\bibinfo{author}{Bertschinger, N.} \& \bibinfo{author}{Natschl\"ager, T.}
\newblock \bibinfo{journal}{\bibinfo{title}{Real-{Time} {Computation} at the
  {Edge} of {Chaos} in {Recurrent} {Neural} {Networks}}}.
\newblock {\emph{\JournalTitle{Neural. Comput.}}}
  \textbf{\bibinfo{volume}{16}}, \bibinfo{pages}{1413} (\bibinfo{year}{2004}).

\bibitem{taniguchi20AIP}
\bibinfo{author}{Taniguchi, T.}
\newblock \bibinfo{journal}{\bibinfo{title}{Synchronization and chaos in spin
  torque oscillator with two free layers}}.
\newblock {\emph{\JournalTitle{AIP Adv.}}} \textbf{\bibinfo{volume}{10}},
  \bibinfo{pages}{015112} (\bibinfo{year}{2020}).

\end{thebibliography}



\section*{Acknowledgements}

T.T. acknowledges Takayuki Nozaki, Tomohiro Nozaki, and Yoichi Shiota for their valuable discussions. 
This paper was based on the results obtained from a project (Innovative AI Chips and Next-Generation Computing Technology Development/(2) 
Development of next-generation computing technologies/Exploration of Neuromorphic Dynamics towards Future Symbiotic Society) commissioned by NEDO. 
The work is also supported by JPS KAKENHI Grant Number 20H05655. 


\section*{Author contributions statement}

T.T. designed the project with help from S.T. 
A.O., Y.U. and T.T. developed the codes and performed the simulations. 
T.T. wrote the manuscript and prepared the figures. 
All authors contributed to discussing the results. 


\section*{Competing interests}

The authors declare no competing interests. 


\section*{Data availability}

The datasets used and/or analysed during the current study available from the corresponding author on reasonable request.


\section*{Additional information}

\textbf{Correspondence} and requests for materials should be addressed to T.T.





\end{document}